\documentclass[aip,jap,reprint]{revtex4-1}
\usepackage[english]{babel}
\usepackage[utf8]{inputenc}
\usepackage[dvips]{graphicx}
\usepackage{fancyhdr}
\usepackage[active]{srcltx}
\usepackage{setspace}
\usepackage{float}
\usepackage{amsmath}
\usepackage{amsfonts}
\begin{document}

\title{Interband polarized absorption in InP polytypic superlattices}

\author{P. E. Faria Junior}
\affiliation{Instituto de F\'isica de S\~ao Carlos, Universidade de S\~ao Paulo, 13566-590 S\~ao Carlos, S\~ao Paulo, Brazil}
\affiliation{Department of Physics, State University of New York at Buffalo, Buffalo, New York 14260, USA}

\author{T. Campos}
\affiliation{Instituto de F\'isica de S\~ao Carlos, Universidade de S\~ao Paulo, 13566-590 S\~ao Carlos, S\~ao Paulo, Brazil}

\author{G. M. Sipahi}
\affiliation{Instituto de F\'isica de S\~ao Carlos, Universidade de S\~ao Paulo, 13566-590 S\~ao Carlos, S\~ao Paulo, Brazil}
\affiliation{Department of Physics, State University of New York at Buffalo, Buffalo, New York 14260, USA}  

%===============================================================================

\begin{abstract}

Recent advances in growth techniques have allowed the fabrication of semiconductor 
nanostructures with mixed wurtzite/zinc-blende crystal phases. Although the optical 
characterization of these polytypic structures is well reported in the literature, 
a deeper theoretical understanding of how crystal phase mixing and quantum confinement 
change the output linear light polarization is still needed. In this paper, we 
theoretically investigate the mixing effects of wurtzite and zinc-blende phases 
on the interband absorption and in the degree of light polarization of an InP 
polytypic superlattice. We use a single 8$\times$8 k$\cdot$p Hamiltonian that 
describes both crystal phases. Quantum confinement is investigated by changing 
the size of the polytypic unit cell. We also include the optical confinement effect 
due to the dielectric mismatch between the superlattice and the vaccum and we show 
it to be necessary to match experimental results. Our calculations for large wurtzite 
concentrations and small quantum confinement explain the optical trends of recent 
photoluminescence excitation measurements. Furthermore, we find a high sensitivity 
to zinc-blende concentrations in the degree of linear polarization. This sensitivity 
can be reduced by increasing quantum confinement. In conclusion, our theoretical 
analysis provides an explanation for optical trends in InP polytypic superlattices, 
and shows that the interplay of crystal phase mixing and quantum confinement is 
an area worth exploring for light polarization engineering.

\end{abstract}

\maketitle

%===============================================================================

\section{Introduction}

The past few years has seen tremendous advances in growth techniques of low dimensional 
semiconductor nanostructures, especially concerning III-V nanowires (NWs). At the 
moment, it is possible to precisely tune the growth conditions to achieve single 
crystal phase nanostructures\cite{Mohan2005,Vu2013,Pan2014} or polytypic heterostructures 
with sharp interfaces\cite{Lehmann2013,Khranovskyy2013}. Moreover, it has been 
reported successful integration of III-V NWs with silicon\cite{Ren2013,Borg2014,Heiss2014,Li2014a}, 
increasing the possibilities for developing new optoelectronic devices\cite{Akopian2010,Smit2014}.

Because of its lower surface recombination and higher electron mobility\cite{Joyce2013,Ponseca2014}, 
InP is a good candidate, among the III-V compounds, to be embedded in these novel devices.
Polytypic InP homojunctions showing a type-II band alignment\cite{Murayama1994} can be explored 
to engineer light polarization\cite{Hoang2010} and to enhance the lifetime of carriers\cite{Pemasiri2009,Yong2013}. 
In fact, the use of InP NWs has been proposed in FETs\cite{Duan2001,Jiang2007,Wallentin2012}, 
silicon integrated nanolasers\cite{Wang2013} and stacked p-n junctions in solar 
cells\cite{Wallentin2013,Cui2013}.

Although the process underlying the formation of these polytypic homojunctions 
is elucidated\cite{Akiyama2006,Kitauchi2010,Ikejiri2011,Poole2012} and an extensive literature 
on the optical characterization of these structures is available\cite{Mattila2006,
Mishra2007,Paiman2009,Gadret2010,Kailuweit2012,Li2014a}, we lack theoretical 
understanding of how the crystal phase mixture changes the light polarization on 
these nanostructures.

\begin{figure}[h!]
\begin{center}
\includegraphics{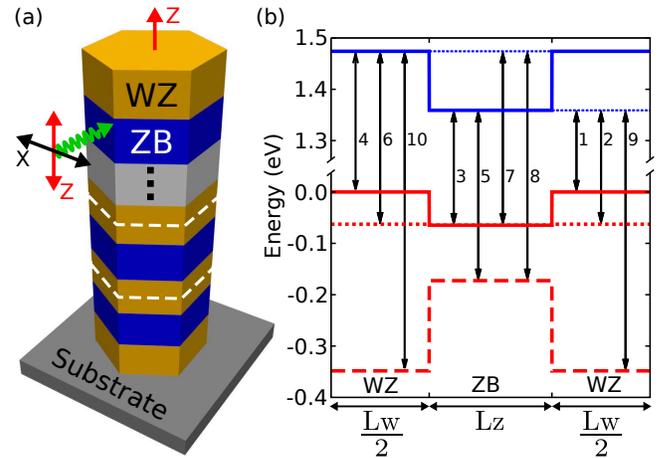}
\caption{(a) InP polytypic superlattice grown along WZ[0001]/ZB[111] direction. 
The (red) arrow on top indicates the growth direction. Polarization of incident 
light can be in X- or Z-direction, i.e., perpendicular or parallel to the growth 
direction. The polytypic superlattice unit cell, or simply supercell, of size 
$L$=Lw+Lz is bounded by dashed (white) lines. (b) Band-edge energy diagram 
at $\vec{k}=0$ of type-II InP supercell with possible transitions. The numbers on 
the side of the vertical arrows indicate the magnitude of energy transitions, i.e., 
1 means the lowest and 10 the highest. For the valence band, solid line is heavy 
hole band for both WZ/ZB, dotted line is light hole band for both WZ/ZB and dashed 
line is crystal-field split-off hole band for WZ and split-off hole band for ZB. 
The small dotted lines in the conduction band are plotted just to guide the eyes.}
\label{fig1}
\end{center}
\end{figure}

The aim of this study is to provide a comprehensive analysis on how wurtzite 
(WZ)/zinc-blende (ZB) mixing, quantum confinement (QC), and also optical confinement 
(OC) modify the interband absorption and the degree of linear polarization (DLP) 
in an InP polytypic superlattice. From now on, we will use the term superlattice 
for the polytypic case. In our calculations, the QC along growth direction takes 
into account the changes of WZ and ZB phases. Also, assuming large crosssections, 
we neglect lateral QC.

A scheme of the superlattice and possible light polarizations is presented in Fig. \ref{fig1}(a). 
Although we show a NW with multiple WZ and ZB segments, the periodicity of these 
segments allows us to consider only the unit cell, bounded by dashed lines, to 
understand the physics of the superlattice. The incoming light polarization can 
be either parallel (Z) or perpendicular (X) to the growth direction.

To calculate the band structure, we extend our polytypic k$\cdot$p method\cite{fariajunior-jap2012} 
and include the conduction and valence band interaction explicitly. This interaction 
increases the reliability of the method, and allows us to calculate the band 
structure further away from the center of the Brillouin zone. Furthermore, we 
provide the parameter sets for WZ and ZB InP in this new 8$\times$8 k$\cdot$p 
configuration. 

We find that the trends of recent photoluminescence (PL) and excitation photoluminescence 
(PLE) measurements performed by Gadret et al.\cite{Gadret2010} can be explained by 
our model. Although their samples are disordered, i.e., the regions of WZ/ZB are 
not periodically ordered, we can predict the observed trends considering a supercell 
of 100 nm composed of 95\% WZ. In addition, we show that the DLP can be tuned using 
WZ/ZB mixing and QC. The limiting cases of our superlattice, i.e., pure ZB and 
pure WZ NWs are also calculated and their DLP around gap energy is in very good 
agreement with the results from Mishra et al. \cite{Mishra2007}. This matching 
emphasizes the use of OC in our calculations.

The structure of the present paper is the following: in Sec. II, we describe the 
8$\times$8 polytypic k$\cdot$p method and present our approach for the interband 
transitions. Section III contains our results for interband absorption and DLP 
in the bulk and superlattice regimes. Finally, in Sec. IV, we summarize our main 
findings and present our conclusions.

%===============================================================================

\section{Theoretical background}

\subsection{Hamiltonian}

We expand the Hamiltonian of Ref.\cite{fariajunior-jap2012} to explicitly include 
the interband interaction. Since there is no coupling between the ZB irreducible 
representations for conduction ($\Gamma_1\sim x^2 + y^2 + z^2$) and valence ($\Gamma_{15} 
\sim x,y,z$) bands, we can apply the same rotation\cite{Park2000} for the [001] 
k$\cdot$p matrix with interband interaction. The total rotation matrix would be 
the direct sum of valence and conduction band rotation matrices, therefore an 
8$\times$8 matrix with 6$\times$6 and 2$\times$2 blocks. An alternative procedure 
would be to start with the Hamiltonian in the [111] coordinate system without 
interband interaction and derive the interband matrix elements in the [111] 
coordinate system relating them to the [001].

Our bulk Hamiltonian basis set, defined at $\Gamma$-point, in the ZB[111]/WZ[0001] 
coordinate system is:

\begin{eqnarray}
\left|c_{1}\right\rangle & = & -\frac{1}{\sqrt{2}}\left|(X+iY)\uparrow\right\rangle \nonumber \\
\left|c_{2}\right\rangle & = & \frac{1}{\sqrt{2}}\left|(X-iY)\uparrow\right\rangle \nonumber \\
\left|c_{3}\right\rangle & = & \left|Z\uparrow\right\rangle \nonumber \\
\left|c_{4}\right\rangle & = & \frac{1}{\sqrt{2}}\left|(X-iY)\downarrow\right\rangle \nonumber \\
\left|c_{5}\right\rangle & = & -\frac{1}{\sqrt{2}}\left|(X+iY)\downarrow\right\rangle \nonumber \\
\left|c_{6}\right\rangle & = & \left|Z\downarrow\right\rangle \nonumber \\
\left|c_{7}\right\rangle & = & i\left|S\uparrow\right\rangle \nonumber \\
\left|c_{8}\right\rangle & = & i\left|S\downarrow\right\rangle 
\label{eq:poly_InP:cbasis}
\end{eqnarray}
\\
with 1-6 representing the valence band states and 7-8 the conduction band states. 
In this basis set, the Hamiltonian including interband interactions is given by

\begin{equation}
\mathbb{H}_{8}=\left[\begin{array}{cc}
\mathbb{H}_\textrm{V} & \mathbb{H}_\textrm{VC} \\
\mathbb{H}^\dagger_\textrm{VC} & \mathbb{H}_\textrm{C}
\end{array}\right]
\label{eq:hamil}
\end{equation}
\\
where $\mathbb{H}_\textrm{V}$ represent the valence band, $\mathbb{H}_\textrm{C}$ 
the conduction band and $\mathbb{H}_\textrm{VC}$ the interaction term between them. 
The sub-matrices have the following forms:

\begin{equation}
\mathbb{H}_\textrm{V}=\left[\begin{array}{ccccccc}
F  & -K^{*} &  -H^{*}  & 0 & 0 & 0 \\
-K &   G    &   H      & 0 & 0 & \sqrt{2}\Delta_{3} \\
-H &  H^{*} &  \lambda & 0 & \sqrt{2}\Delta_{3} & 0 \\
0  & 0      & 0 & F & -K & H \\
0  & 0      & \sqrt{2}\Delta_{3} & -K^{*} & G & -H^{*} \\
0  & \sqrt{2}\Delta_{3} & 0 & H^{*} & -H & \lambda 
\end{array}\right]
\end{equation}

\begin{equation}
\mathbb{H}_\textrm{VC}=\left[\begin{array}{cc}
-\frac{1}{\sqrt{2}} P_2 k_-   & 0 \\
 \frac{1}{\sqrt{2}} P_2 k_+   & 0 \\
                    P_{1} k_z & 0 \\
 0 & \frac{1}{\sqrt{2}} P_2 k_+   \\
 0 & -\frac{1}{\sqrt{2}} P_2 k_- \\
 0 & P_1 k_z
\end{array}\right]
\end{equation}

\begin{equation}
\mathbb{H}_\textrm{C}=\left[\begin{array}{cc}
E_C & 0 \\
0 & E_C
\end{array}\right]
\end{equation}
\\
and their terms

\begin{eqnarray}
F & = & \Delta_{1}+\Delta_{2}+\lambda+\theta\nonumber \\
G & = & \Delta_{1}-\Delta_{2}+\lambda+\theta\nonumber \\
\lambda & = & \tilde{A}_{1}k_{z}^{2}+\tilde{A}_{2}\left(k_{x}^{2}+k_{y}^{2}\right)\nonumber \\
\theta & = & \tilde{A}_{3}k_{z}^{2}+\tilde{A}_{4}\left(k_{x}^{2}+k_{y}^{2}\right)\nonumber \\
K & = & \tilde{A}_{5}k_{+}^{2} + 2\sqrt{2}\tilde{A}_z k_- k_z \nonumber \\
H & = & \tilde{A}_{6}k_{+}k_{z} + \tilde{A}_z k^2_-\nonumber \\
E_{C} & = & E_{g}+E_{0}+\tilde{e}_{1}k_{z}^{2}+\tilde{e}_{2}\left(k_{x}^{2}+k_{y}^{2}\right)
\label{eq:terms}
\end{eqnarray}
\\
where $\tilde{A}_1 ... \tilde{A}_6, \tilde{A}_z$ and $\tilde{e}_{1}, \tilde{e}_{2}$, 
given in units of $\hbar^{2}/2m_{0}$, are the effective mass parameters of 
valence and conduction band, respectively. Here $\Delta_1$ is the crystal field 
splitting energy in WZ, $\Delta_2,\Delta_3$ are the spin-orbit coupling splitting energies, 
$k_{\pm}=k_{x}\pm ik_{y}$ and $P_1,P_2$ are the Kane parameters of the interband 
interactions, given by

\begin{eqnarray}
P_1 & = & -i \frac{\hbar}{m_0} \left\langle X \left| p_{x} \right|S \right\rangle = -i \frac{\hbar}{m_0} \left\langle Y \left| p_{y} \right| S \right\rangle \nonumber \\
P_2 & = & -i \frac{\hbar}{m_0} \left\langle Z \left| p_{z} \right| S \right\rangle 
\end{eqnarray}

We would like to emphasize that the Hamiltonian (\ref{eq:hamil}) and its terms 
(\ref{eq:terms}) describe both WZ and ZB crystal structures, however, the usual 
ZB parameters must be mapped to the ones in equation (\ref{eq:terms}). Moreover, 
the inclusion of the interband interaction explicitly in the Hamiltonian also 
requires some corrections to be made in the second order effective mass parameters. 
These corrections appear because conduction and valence band states are now treated 
as belonging to the same class, following Löwdin's notation \cite{lowdin-jcp1951}. 
We describe the mapping and corrections of parameters with detail in Appendix \ref{app:2nd_corr}.

In order to treat the confined direction along $z$, we apply the envelope function 
approximation\cite{Bastard1981,Baraff1991} to the Hamiltonian (\ref{eq:hamil}). 
This treatment can be summarized with the following changes:

\begin{eqnarray}
g & \rightarrow & g(z) \nonumber \\
k_z & \rightarrow & -i\frac{\partial}{\partial z} \nonumber \\
Bk^2_z & \rightarrow & -\frac{\partial}{\partial z} B(z) \frac{\partial}{\partial z} \nonumber \\
Pk_z & \rightarrow & - \frac{i}{2} \left[ \frac{\partial}{\partial z}P(z) + P(z)\frac{\partial}{\partial z} \right]
\end{eqnarray}
\\
with $g$ representing the parameters in the Hamiltonian (different in WZ and ZB), 
$B$ representing an effective mass parameter and $P$ is the interband parameter. 
The last two equations are the symmetrization requirements to hold the Hermitian 
property of the momentum operator\cite{Baraff1991}. Any parameter in the Hamiltonian 
acquires a dependence along $z$, making it different for WZ and ZB. Also, the 
confinement profile due to the polytypic interface is added to the Hamiltonian. 
In Fig. \ref{fig1}(b), we present the InP band-edge profile along $z$ for $\vec{k}=0$, 
which takes into account the interface profile and the intrinsic splittings of 
each crystal structure.

Under the envelope function approximation, a general state $n,\vec{k}$ of the 
superlattice can be written as

\begin{equation}
\psi_{n,\vec{k}}\left(\vec{r}\right)=e^{i\vec{k}\cdot\vec{r}}\sum_{m=1}^{8}f_{n,\vec{k},m}(z) \, u_{m}(\vec{r})
\end{equation}

We then apply the plane wave expansion for the parameters and the envelope functions
to solve the Hamiltonian numerically. Since this expansion automatically considers 
periodic boundary conditions, we can associate the value $k_z$ for the superlattice, 
$-\frac{\pi}{L} \leq k_z \leq \frac{\pi}{L}$, while the Fourier coefficients have 
$K_j = j\frac{2\pi}{L}$ (with $j=0,\pm1,\pm2,\ldots$). For confined states, the 
dispersion of the band structure along $k_z$ is a flat band. However, higher energy 
states are no longer confined and does not have this flat dispersion. Since we are 
interested in transitions that also take into account these higher energy states, 
we will include the $k_z$ in our calculation.

%-------------------------------------------------------------------------------

\subsection{Interband absorption}

The absorption coefficient\cite{book-chuang} of photons with energy $\hbar\omega$ 
can be written as

\begin{equation}
\alpha_{\hat{\epsilon}}(\hbar\omega)=\frac{C}{\hbar\omega}\underset{a,b,\vec{k}}{\sum}\, I_{a,b,\vec{k}}^{\hat{\epsilon}}\,\left(f_{a,\vec{k}}-f_{b,\vec{k}}\right)\,\mathcal{L}^\Gamma_{a,b,\vec{k}}(\hbar\omega)
\label{eq:poly_InP:absorption}
\end{equation}
\\
where $a$ ($b$) runs over conduction (valence) sub-bands, $\vec{k}$ runs over reciprocal 
space points, $\hat{\epsilon}$ is the light polarization, $I^{\hat{e}}_{a,b,\vec{k}}$ 
is the interband dipole transition amplitude, $\mathcal{L}^\Gamma_{a,b,\vec{k}}(\hbar\omega)$ 
gives the transition broadening and $f_{a(b),\vec{k}}$ is the Fermi-Dirac 
distribution of conduction (valence) band. We will consider $T=0\;\text{K}$ and 
no doping, i.e., the valence band is full and the conduction band is empty\cite{prb-82-235425}, 
leading to $f_{a,\vec{k}}-f_{b,\vec{k}}=1$. The constant $C$ is given by

\begin{equation}
C = \frac{\pi\hbar e^{2}}{cn_{r}\varepsilon_{0}m_{0}^{2}\Omega}
\end{equation}
\\
where $e$ is the electron charge, $c$ is the velocity of light, $n_r$ is the 
refractive index of the material, $\epsilon_0$ is the vacuum dielectric constant, 
$m_0$ is the free electron mass, and $\Omega$ is the volume of the real space.

We considered a Lorentzian broadening for the transitions

\begin{equation}
\mathcal{L}^\Gamma_{a,b,\vec{k}}(\hbar\omega) = \frac{1}{2\pi}\frac{\Gamma}{\left[E_{a}(\vec{k}) - E_{b}(\vec{k}) -\hbar\omega\right]^{2}+\left(\frac{\Gamma}{2}\right)^{2}}
\end{equation}
\\
with $\Gamma$ as the full width at half-maximum. In our calculations, we set 
$\Gamma=2\;\text{meV}$.

For $\hat{x}$ and $\hat{z}$ light polarizations, the interband dipole transition 
amplitudes, between conduction ($a$) and valence ($b$) states, are given by

\begin{eqnarray}
\ensuremath{I_{a,b,\vec{k}}^{\hat{x}}} & \propto & \frac{1}{2}\left| \left\langle F_{a,\vec{k},7}|F_{b,\vec{k},1}\right\rangle +\left\langle F_{a,\vec{k},1}|F_{b,\vec{k},7}\right\rangle \right.\nonumber \\
 &  & \;-\left\langle F_{a,\vec{k},7}|F_{b,\vec{k},2}\right\rangle -\left\langle F_{a,\vec{k},2}|F_{b,\vec{k},7}\right\rangle \nonumber \\
 &  & \;-\left\langle F_{a,\vec{k},8}|F_{b,\vec{k},4}\right\rangle -\left\langle F_{a,\vec{k},4}|F_{b,\vec{k},8}\right\rangle \nonumber \\
 &  & \;+\left.\left\langle F_{a,\vec{k},8}|F_{b,\vec{k},5}\right\rangle +\left\langle F_{a,\vec{k},5}|F_{b,\vec{k},8}\right\rangle \right|^{2}
\label{eq:dipole_x}
\end{eqnarray}

\begin{eqnarray}
\ensuremath{I_{a,b,\vec{k}}^{\hat{z}}} & \propto & \left|\left\langle F_{a,\vec{k},7}|F_{b,\vec{k},3}\right\rangle +\left\langle F_{a,\vec{k},3}|F_{b,\vec{k},7}\right\rangle \right.\nonumber \\
 &  & \!\!\!\!+\left.\left\langle F_{a,\vec{k},8}|F_{b,\vec{k},6}\right\rangle +\left\langle F_{a,\vec{k},6}|F_{b,\vec{k},8}\right\rangle \right|^{2}
\label{eq:dipole_z}
\end{eqnarray}
\\
with
\begin{equation}
\left\langle F_{a,\vec{k},m}|F_{b,\vec{k},n}\right\rangle = \frac{1}{L} \int_{L}dz\, f_{a,\vec{k},m}^{*}(z)\,f_{b,\vec{k},n}(z)
\end{equation}
\\
where $L$ is the size of the supercell and the factor $1/L$ appears because 
our envelope functions are normalized in reciprocal space.

We assumed the interband coupling parameters to be constant throughout the polytypic 
system, i.e., the same values were used for both polytypes since their numerical 
values do not differ much (see Table \ref{tab:kp_par}).

The relative contributions of  the different light polarizations can be probed 
by analyzing the DLP:

\begin{equation}
\text{DLP}(\hbar\omega)=\frac{\alpha_{\hat{z}}(\hbar\omega)-\alpha_{\hat{x}}(\hbar\omega)}{\alpha_{\hat{z}}(\hbar\omega)+\alpha_{\hat{x}}(\hbar\omega)}
\label{eq:DLP}
\end{equation}
\\ which ranges from $-1$, if the absorbed light is polarized perpendicular to the 
wire axis, to $1$, if polarization is parallel to the growth direction.

We have also investigated the effect of OC due to the dielectric mismatch between 
the vacuum and the superlattice. This effect is included as follows\cite{PhysRevB.70.165317}:

\begin{equation}
\alpha_{\hat{x}}(\hbar\omega) \rightarrow \frac{2}{1 + \epsilon} \alpha_{\hat{x}}(\hbar\omega)
\label{eq:OCeffects}
\end{equation}
\\
with $\epsilon$ being the dielectric constant of the superlattice, which was considered 
to be the same for ZB and WZ InP ($\epsilon = 12.4$\cite{Wang2001,PhysRevB.70.161310}).

%===============================================================================

\section{Results and discussion}

\subsection{Bulk}

\begin{table}[H]
\begin{center}
\caption{k$\cdot$p parameters of the polytypic 8$\times$8 model for InP.}
\begin{tabular}{ccc}
\hline 
\hline 
Parameter & ZB InP & WZ InP \tabularnewline
\hline 

Lattice constant ($\textrm{\AA}$) &  & \tabularnewline

$a$ & 4.1505  & 4.1505 \tabularnewline
$c$ & 10.1666 & 6.7777 \tabularnewline

Energy parameters (eV) &  & \tabularnewline

$E_{g}$ & 1.4236 & 1.474 \tabularnewline

$\Delta_{1}$            & 0     & 0.303 \tabularnewline
$\Delta_{2}=\Delta_{3}$ & 0.036 & 0.036 \tabularnewline

Conduction band effective mass \\ parameters (units of $\frac{\hbar^{2}}{2m_{0}}$) &  & \tabularnewline

$\tilde{e}_{1}$ & -1.6202 & -1.2486 \tabularnewline
$\tilde{e}_{2}$ & -1.6202 & -1.6231 \tabularnewline

Valence band effective mass \\ parameters (units of $\frac{\hbar^{2}}{2m_{0}}$) &  & \tabularnewline

$\tilde{A}_{1}$ &  1.0605 &  0.0568 \tabularnewline
$\tilde{A}_{2}$ & -0.8799 & -0.8299 \tabularnewline
$\tilde{A}_{3}$ & -1.9404 & -0.8423 \tabularnewline
$\tilde{A}_{4}$ &  0.9702 &  1.2001 \tabularnewline
$\tilde{A}_{5}$ &  1.4702 & 11.4934 \tabularnewline
$\tilde{A}_{6}$ &  2.7863 &  9.8272 \tabularnewline
$\tilde{A}_{z}$ & -0.5000 &  0      \tabularnewline

Interband coupling parameters (eV\AA) &  & \tabularnewline
$P_1=P_2$ & 8.7249 & 8.3902 \tabularnewline

\hline 
\hline 
\label{tab:kp_par}
\end{tabular}
\end{center}
\end{table}

Before we turn to the superlattice case, it is useful to understand the light polarization 
properties for bulk ZB and WZ. Indeed, these would be the limiting cases of our 
superlattice calculations. We can view these bulk limiting cases as NWs of large 
diameter and length, with pure crystal phases. Here, we also assumed the light 
polarizations described in Fig. \ref{fig1}(a). Moreover, the 8$\times$8 parameter 
sets were derived from the 6$\times$6 model of our previous paper \cite{fariajunior-jap2012}, which 
was based on the effctive masses of Ref.\cite{prb-81-155210}. The lattice constants 
$a$ and $c$ of ZB are given in the [111] unit cell (ZB has 3 bilayers of atoms 
instead of 2 in WZ). Table \ref{tab:kp_par} have all the parameters used in the 
simulations.

\begin{figure}[h!]
\begin{center}
\includegraphics{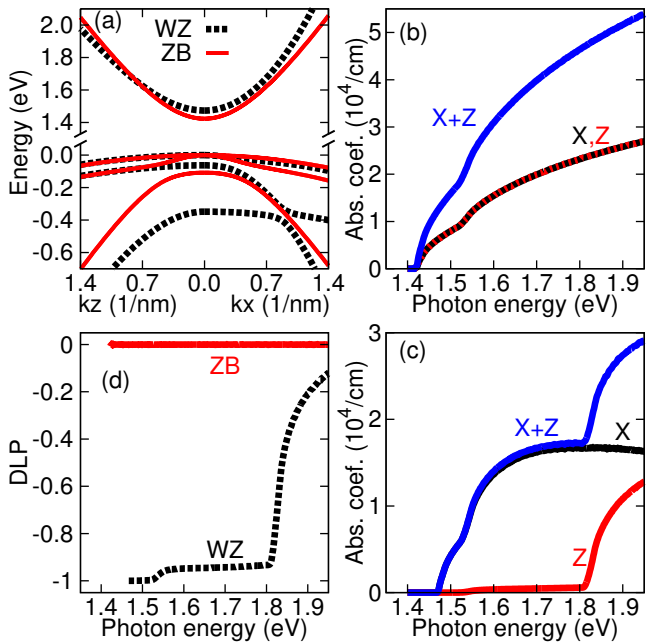}
\caption{In clockwise order: (a) InP bulk band structure for WZ (dashed lines) 
and ZB (solid lines). InP Bulk ZB (b) and WZ (c) absorption coefficients as 
functions of the photon energies. (d) ZB (solid line) and WZ (dashed line) DLP. 
The crystal structure plays an important role in the absorption: ZB absorbs light 
isotropically while WZ have a clear anisotropy that changes when photon energy 
reaches the band edge energies.}
\label{fig2}
\end{center}
\end{figure}

In Fig. \ref{fig2}(a) we show the band structure of bulk ZB[111] and WZ[0001] for 
$k_x$ and $k_z$ directions. At $\vec{k}=0$, the valence band of WZ has three 
energy bands two-fold degenerate, while ZB has one four-fold degenerate band and 
one two-fold degenerate. From top to bottom, WZ valence bands are labeled HH (heavy hole), 
LH (light hole) and CH (crystal-field split-off hole), following Chuang and Chang's 
notation\cite{apl-68-1657}, and ZB valence bands are labeled usually as HH/LH and SO 
(split-off hole). Each band-edge in the band structure will have a signature in 
the absorption spectra, therefore, we can expect three regions for WZ and only 
two for ZB. Moreover, the symmetry of the eigenstates will rule the light polarization, 
as shown by equations (\ref{eq:dipole_x}) and (\ref{eq:dipole_z}).

Figures \ref{fig2}(b) and \ref{fig2}(c) show the bulk absorption coefficients for 
ZB and WZ, respectively, as calculated by equation (\ref{eq:poly_InP:absorption}).
Although we considered ZB in the [111] unit cell, X- and Z-polarizations have the 
same absorption as we would expect from the cubic symmetry. One can easily show 
that the coordinate system rotation we have applied holds this cubic character in 
the absorption coefficient. Note the shoulder in the curve when the photon 
energy reaches the SO band energy. For WZ, however, a clear anisotropy between X- 
and Z-polarizations is found. Before we reach the CH band energy, light is 
predominantly X-polarized, however, after CH band the Z-polarized absorption 
increases while X-polarized slightly decreases.

To highlight the polarization differences for ZB and WZ, we show in Fig. \ref{fig2}(d) 
the DLP, given by equation (\ref{eq:DLP}). Since X- and Z-polarizations are the 
same in ZB, we have a straight line at DLP=0, meaning isotropic absorption. For 
WZ, the DLP starts at $-1$, slightly increases when LH band is reached and after CH 
band, it rapidly goes to 0 due to the Z-polarization contribution. In the superlattice 
calculations, we expect that the WZ/ZB mixing and QC effects will change the DLP 
to some intermediate value between pure ZB and WZ.

%-------------------------------------------------------------------------------

\subsection{Absorption and PLE measurements}

\begin{figure}[h!]
\begin{center}
\includegraphics{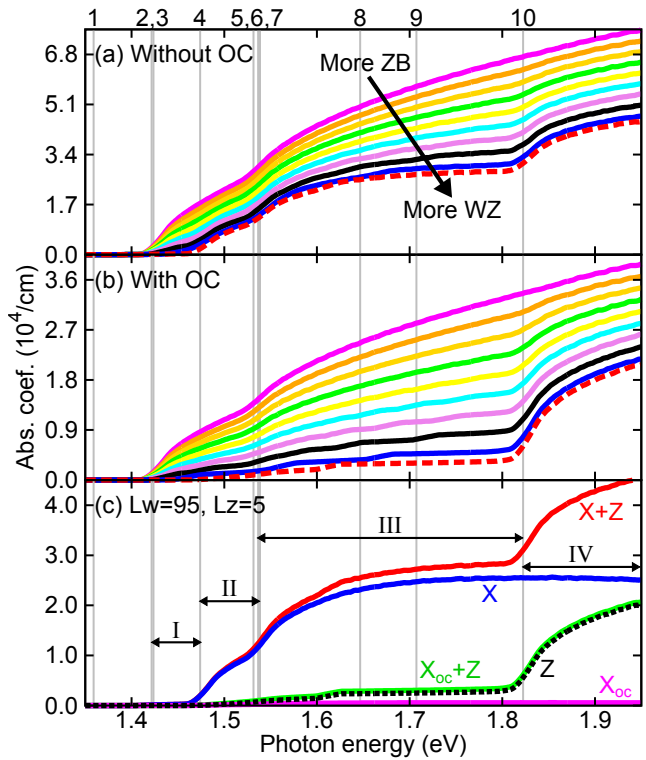}
\caption{Absorption coefficient for 100 nm supercell with different 
WZ/ZB ratio (a) without OC and (b) with OC. From top to bottom, solid lines range 
from 10\%WZ/90\%ZB to 90\%WZ/10\%ZB with steps of 10\%. The dashed line is the 
95\%WZ/5\%ZB regime. (c) Absorption coefficient for 95\%WZ/5\%ZB, including the 
contributions for X- and Z-polarizations. The numbers at the top of the Fig. 
indicate the transition energies of Fig. \ref{fig1}(b).}
\label{fig3}
\end{center}
\end{figure}

Let us start the superlattice investigation by considering small QC, i.e., relatively 
large WZ and ZB regions (5-90 nm each with total supercell of 100 nm), values 
typically found in superlattice samples. Although there is small QC, small regions 
of WZ and ZB act as perturbations to the bulk states leading to different electronic 
and optical properties.

In Fig. \ref{fig3}(a) we show the total absorption, $\alpha_{\hat{x}}+\alpha_{\hat{z}}$, 
without OC effects. The first (top) solid curve is the case of 90\% of ZB and 10\% 
of WZ and we can notice the two characteristic shoulders of the bulk case, the first
around transition energy 3 and the second around transition 5. Increasing the mixing 
of ZB and WZ, new shoulders appear and when we reach the last (bottom) solid curve, 
90\% of WZ and 10\% of ZB, we notice the three characteristic shoulders of WZ bulk 
case, around transitions 4, 6 and 10, respectively. This WZ characteristic is also 
noted in the dashed line, which is the 95\%WZ/5\%ZB. We can also notice from Fig. \ref{fig3}(a) 
that there is a non-zero absorption coefficient between ZB and WZ energy gaps (transitions 
2 and 4) even for large WZ concentrations. This phenomenon is due to the characteristic 
type-II band alignment of ZB/WZ homojunctions.

\begin{figure}[h!]
\begin{center}
\includegraphics{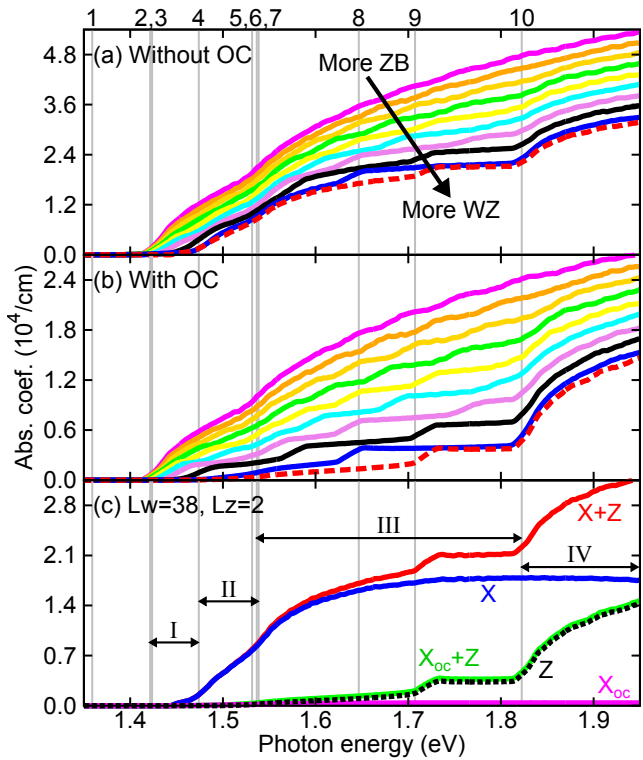}
\caption{Same as system Fig. 3 for 40 nm supercell. In this case, because of the 
smaller supercell, QC effects are more pronounced.}
\label{fig4}
\end{center}
\end{figure}

When we add OC effects, which are presented in Fig. \ref{fig3}(b), we notice 
a significant suppression of X-polarization that becomes more evident as WZ/ZB ratio 
increases. Since in WZ the absorption spectrum in regions I and II mainly 
comprises sub-bands with a mixture of states $\left|c_{1}\right\rangle$, $\left|c_{4}\right\rangle$ 
due to HH symmetry, the OC almost forbids the X component from penetrate the NW, 
therefore the suppression.

The experimental paper of Gadret et al.\cite{Gadret2010} investigates optical 
properties of InP polytypic superlattices. They report PL and PLE measurements 
of InP polytypic samples with statistically negligible percentage of ZB. In this 
regime, they notice 3 absorption edges in the PLE ($\sim$1.488 eV, $\sim$1.532 eV 
and $\sim$1.675 eV) for energies above the PL peak ($\sim$1.432 eV) and also a long 
tail at the low energy side of the PL peak. Their system is comparable to our simulation 
for 100 nm supercell with 95\% of WZ and 5\% of ZB or even higher WZ percentage 
over ZB. Indeed, the measured trends are well described by our Fig. \ref{fig3}(c). 
We can identify 4 regions that we can relate to the experimental spectra: I (between 
transitions 2/3 and 4, i.e., between ZB and WZ gap energies), II (between transitions 
4 and 5/6/7), III (between transitions 5/6/7 and 10) and IV (after transition 10). 
From the observed data we can assign the three absorption edges to the the beginning 
of energy regions II, III and IV, respectively. Region I actually comprises the 
region where the PL peak is observed. Furthermore, the long tail at low energy side 
of PL can be explained by the type-II confinement of WZ/ZB interface, which has 
negligible absorption coefficient. Since we are not considering excitons, we do not 
observe the peaks in the absorption at the band-edge transitions (visible in the 
experimental data). However, the band-edge character is well described by our model, 
represented by the shoulders in our graphs. The blue-shift of our band-edge transition 
energies compared to the experimental data is also related to the lack of excitonic 
effects in our model.

For comparison, we plotted in Fig. \ref{fig4} the same results but for a 40 nm 
supercell. In Figs. \ref{fig4}(a) and \ref{fig4}(b) we observe the same trends 
as before but with more quantization effects, signalled by the extra shoulders or 
step-like behavior in absorption spectra. As we increase the QC, the number of 
sub-bands in the same energy range decreases, leading to clear shoulders in the 
spectra as the photon energy reaches these few sub-bands. In Fig. \ref{fig4}(c), 
we show the different contribution of X- and Z-polarizations. Comparing to Fig. 
\ref{fig3}(c), it can be seen that the QC effect is more visible in Z-polarization 
since this is the confined direction. For the X-polarization, a small red-shift 
is observed due to greater overlap between states.

%-------------------------------------------------------------------------------

\subsection{Quantum confinement and crystal phase mixing effects in the DLP}

For a better understanding of the optical properties of the InP homojunctions
it is valuable to study the DLP. Specifically, we are interested in how polarization
properties are modified by different crystal phase mixing and QC.

\begin{figure}[h!]
\begin{center}
\includegraphics{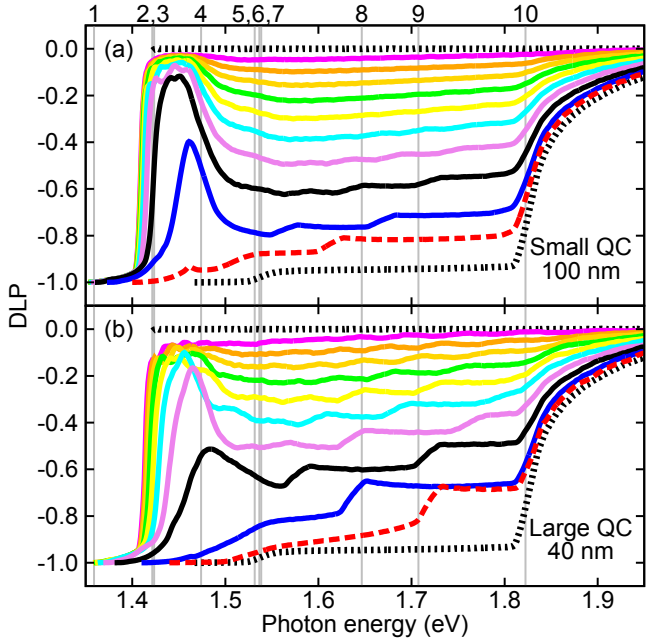}
\caption{Degree of linear polarization without optical confinement for small QC 
(a) and large QC (b). The straight dotted line at 0 indicate the bulk ZB limit 
whereas the lowest dotted line is the bulk WZ limit. The solid and dashed lines 
have the same meaning as in Figs. \ref{fig3} and \ref{fig4}.}
\label{fig5}
\end{center}
\end{figure}

In Fig. \ref{fig5} we show the behavior of the DLP only under QC effects, no OC 
included, for the case in which the supercell has 100 nm, Fig. \ref{fig5}(a), 
and 40 nm, Fig. \ref{fig5}(b). In general, the DLP is very sensitive to ZB insertions:
in region I it is close to 0, exception made to systems where WZ regions are largely 
dominant over ZB ones, about 80\%WZ or more for the small QC regime and 70\%WZ or 
more in large QC. For all different WZ/ZB mixing, the limits are bulk WZ and bulk 
ZB DLP, presented in Fig. \ref{fig2}(d) and showed here with dotted lines.

\begin{figure}[h!]
\begin{center}
\includegraphics{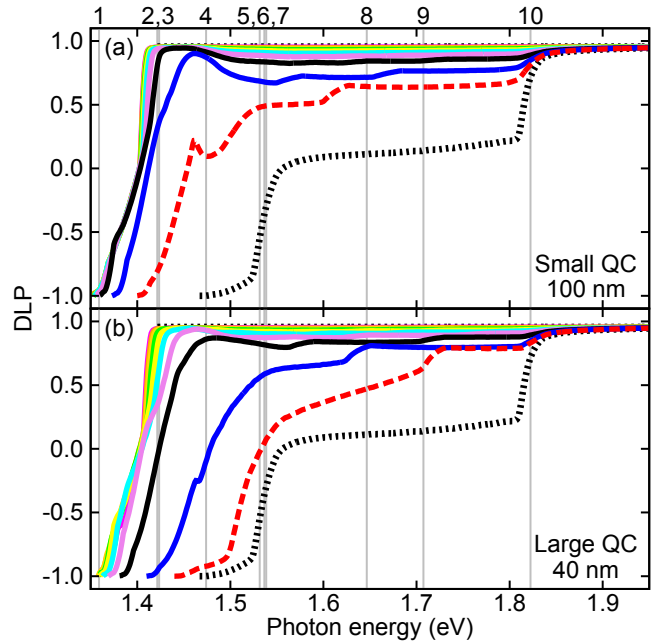}
\caption{Same as Fig. \ref{fig5} but including optical confinement. The straight 
dotted line at 1 indicates the bulk ZB limit whereas the lowest dotted line is 
the bulk WZ limit.}
\label{fig6}
\end{center}
\end{figure}

To further analyze the effect of the OC, we present in Fig. \ref{fig6} the DLP 
calculations for the same systems previously discussed, including QC and OC effects. 
Here, we also include OC effects in bulk calculations. In the paper of Mishra et al. \cite{Mishra2007}, 
their PL measurements for pure WZ and ZB NWs with large diameters indicate that for 
ZB light is strongly polarized along the NW axis, whereas for WZ light is strongly 
polarized in the perpendicular direction. Our calculations for the bulk case with 
OC are in very good agreement with these experimental results. In fact, this 
indicates that OC is a necessary feature to be included in the description. Also, 
these are the limiting cases for all WZ/ZB mixed systems.

Comparing the results with OC for small QC, Fig. \ref{fig6}(a), and large QC, 
Fig. \ref{fig6}(b), we also notice that the DLP is very susceptible to ZB concentration, 
i.e., just a small amount of ZB can switch the DLP from -1 to approximately 1. 
Moreover, for large QC, WZ features hold more effectively around the gap energy. 
This happens because WZ holes (parallel to growth direction) have larger effective 
masses ($m^{\text{ZB}}_{HH} = 0.532$ and $m^{\text{WZ}}_{HH\parallel} = 1.273$) and, 
therefore, are the dominant symmetry to light polarization. Hence, increasing QC 
can reduce the ZB susceptibility to the DLP.

As a final remark, if we compare our DLP for small QC to the PL spectra measurement 
of Gadret et al.\cite{Gadret2010}, we also notice a non-zero value for parallel 
polarization. Since their NWs have a statistically negligible percentage of ZB and 
still present parallel polarization, we believe that this corroborates our results 
of ZB susceptibility to the DLP.

%===============================================================================

\section{Conclusions}

We have expanded our previous polytypic k$\cdot$p model\cite{fariajunior-jap2012} to 
include interband coupling explicitly. The validation of our 8$\times$8 model was 
considered for the bulk case and we found how selection rules for WZ and ZB allow 
different light polarization features.

For the InP polytypic superlattice, we found good agreement between our results 
and the experimental measurements of PLE and light polarization. Although we have 
not considered excitonic effects, the energy regions from the paper of Gadret et. 
al.\cite{Gadret2010} can be mapped to our calculations with small QC and large WZ 
composition. When QC is increased, step-like features are observed in the interband 
absorption for Z-polarization, which is parallel to the growth direction.

Since WZ and ZB present different optical selection rules, any mixing of these two 
crystal phases should combine these different light polarizations. Our DLP calculations 
for pure ZB or pure WZ are in good agreement with experiments of Mishra et al. 
\cite{Mishra2007} and OC effects were necessary to match the experimental results. 
Stronger QC retains WZ behavior of the DLP since WZ HH have larger effective mass than 
ZB HH. In the polytypic cases, we found that the DLP is very susceptible to ZB 
regions and only a small amount of ZB drastically increases the DLP. This ZB 
susceptibility is increased if QC effects are decreased. Furthermore, our results 
for the DLP also explain the polarized PL measured by Gadret et al.\cite{Gadret2010}.

In summary, we believe that our findings provide a theoretical explanation for 
the optical properties observed in InP polytypic superlattices and also indicate 
how linear light polarization can be tuned using QC and crystal phase mixing. We 
wish to emphasize that our theoretical approach is not limited only to InP but also 
could be applied to other III-V compounds that exhibit polytypism\cite{prb-81-155210}.

%===============================================================================

\section{Acknowledgements}

The authors acknowledge financial support from the Brazilian agencies CNPq (grants 
\#138457/2011-5, \#246549/2012-2 and \#149904/2013-4) and FAPESP (grants \#2011/19333-4, 
\#2012/05618-0 and \#2013/23393-8). The authors thank James P. Parry and Karie Friedman 
for kindly proofreading the paper.

%===============================================================================

\appendix

\section{Second-order corrections}
\label{app:2nd_corr}

When the interband coupling is considered in k$\cdot$p Hamiltonians, it is necessary 
to correct some of the second order parameters due to the modification of states 
that belong to classes A and B\cite{lowdin-jcp1951}. For the Luttinger parameters 
in ZB, we have

\begin{eqnarray}
\tilde{\gamma}_{1} & = & \gamma_{1}-\frac{E_{P}}{3E_{g}}\nonumber \\
\tilde{\gamma}_{2} & = & \gamma_{2}-\frac{E_{P}}{6E_{g}}\nonumber \\
\tilde{\gamma}_{3} & = & \gamma_{3}-\frac{E_{P}}{6E_{g}}\nonumber \\
F & = & \frac{1}{m_{e}^{*}}-\frac{E_{g}+\frac{2}{3}\Delta_{SO}}{E_{g}+\Delta_{SO}}\frac{E_{P}}{E_{g}} \nonumber \\
E_{P} & = & \frac{2m_{0}}{\hbar^{2}}P^{2}
\end{eqnarray}
\\
and for WZ parameters, we have

\begin{eqnarray}
\tilde{A}_{1} & = & A_{1}+\frac{E_{P1}}{E_{g}+\Delta_{1}}\nonumber \\
\tilde{A}_{2} & = & A_{2}\nonumber \\
\tilde{A}_{3} & = & A_{3}-\frac{E_{P1}}{E_{g}+\Delta_{1}}\nonumber \\
\tilde{A}_{4} & = & A_{4}+\frac{1}{2}\frac{E_{P2}}{E_{g}}\nonumber \\
\tilde{A}_{5} & = & A_{5}+\frac{1}{2}\frac{E_{P2}}{E_{g}}\nonumber \\
\tilde{A}_{6} & = & A_{6}+\frac{1}{\sqrt{2}}\frac{\sqrt{E_{P1}E_{P2}}}{E_{g}+\frac{\Delta_{1}}{2}}\nonumber \\
\tilde{e}_{1} & = & e_{1}-\frac{E_{P1}}{E_{g}+\Delta_{1}}\nonumber \\
\tilde{e}_{2} & = & e_{2}-\frac{E_{P2}}{E_{g}} \nonumber \\
E_{P1(2)}     & = & \frac{2m_{0}}{\hbar^{2}}P_{1(2)}^{2}
\end{eqnarray}

Given the corrected Luttinger parameters, we only have to connect them to the 
ones in the Hamiltonian (\ref{eq:hamil}), using the same idea as presented in our 
previous paper\cite{fariajunior-jap2012}:

\begin{eqnarray}
\Delta_{1} & = & 0\nonumber \\
\Delta_{2} & = & \Delta_{3}=\frac{\Delta_{SO}}{3}\nonumber \\
\tilde{A}_{1} & = & -\tilde{\gamma}_{1}-4\tilde{\gamma}_{3}\nonumber \\
\tilde{A}_{2} & = & -\tilde{\gamma}_{1}+2\tilde{\gamma}_{3}\nonumber \\
\tilde{A}_{3} & = & 6\tilde{\gamma}_{3}\nonumber \\
\tilde{A}_{4} & = & -3\tilde{\gamma}_{3}\nonumber \\
\tilde{A}_{5} & = & -\tilde{\gamma}_{2}-2\tilde{\gamma}_{3}\nonumber \\
\tilde{A}_{6} & = & -\sqrt{2}\left(2\tilde{\gamma}_{2}+\tilde{\gamma}_{3}\right)\nonumber \\
\tilde{A}_{z} & = & \tilde{\gamma}_{2}-\tilde{\gamma}_{3} \nonumber \\
\tilde{e}_{1} & = & \tilde{e}_{2} = F \nonumber \\
P_1   & = & P_2 = P
\end{eqnarray}

For the numerical values presented in Table \ref{tab:kp_par}, we first corrected 
the ZB parameters and then applied the connection to the polytypic Hamiltonian.

%===============================================================================

\bibliography{references}

\end{document}